\documentclass[preprint]{aastex631}

\usepackage{amsmath} 
\usepackage{mathtools}

\begin{document}

\title{The NANOGrav 15-year Data Set: Search for Gravitational Scattering of Pulsars by Free-Floating Objects in Interstellar Space}

\author[0000-0002-2554-0674]{Lankeswar Dey}
\affiliation{Department of Physics and Astronomy, West Virginia University, P.O. Box 6315, Morgantown, WV 26506, USA}
\affiliation{Center for Gravitational Waves and Cosmology, West Virginia University, Chestnut Ridge Research Building, Morgantown, WV 26505, USA}
\affiliation{Institute of Astrophysics, FORTH, GR-71110, Heraklion, Greece}

\author[0000-0003-1082-2342]{Ross J. Jennings}
\altaffiliation{NANOGrav Physics Frontiers Center Postdoctoral Fellow}
\affiliation{Department of Physics and Astronomy, West Virginia University, P.O. Box 6315, Morgantown, WV 26506, USA}
\affiliation{Center for Gravitational Waves and Cosmology, West Virginia University, Chestnut Ridge Research Building, Morgantown, WV 26505, USA}

\author[0009-0006-8984-9220]{Jackson D. Taylor}
\affiliation{Department of Physics and Astronomy, West Virginia University, P.O. Box 6315, Morgantown, WV 26506, USA}
\affiliation{Center for Gravitational Waves and Cosmology, West Virginia University, Chestnut Ridge Research Building, Morgantown, WV 26505, USA}

\author[0000-0003-4090-9780]{Joseph Glaser}
\affiliation{Department of Physics and Astronomy, West Virginia University, P.O. Box 6315, Morgantown, WV 26506, USA}
\affiliation{Center for Gravitational Waves and Cosmology, West Virginia University, Chestnut Ridge Research Building, Morgantown, WV 26505, USA}

\author[0000-0001-7697-7422]{Maura A. McLaughlin}
\affiliation{Department of Physics and Astronomy, West Virginia University, P.O. Box 6315, Morgantown, WV 26506, USA}
\affiliation{Center for Gravitational Waves and Cosmology, West Virginia University, Chestnut Ridge Research Building, Morgantown, WV 26505, USA}

\author[0000-0001-5134-3925]{Gabriella Agazie}
\affiliation{Center for Gravitation, Cosmology and Astrophysics, Department of Physics and Astronomy, University of Wisconsin-Milwaukee,\\ P.O. Box 413, Milwaukee, WI 53201, USA}

\author[0000-0002-8935-9882]{Akash Anumarlapudi}
\affiliation{Department of Physics and Astronomy, University of North Carolina, Chapel Hill, NC 27599, USA}

\author[0000-0003-0638-3340]{Anne M. Archibald}
\affiliation{Newcastle University, NE1 7RU, UK}

\author[0009-0008-6187-8753]{Zaven Arzoumanian}
\affiliation{X-Ray Astrophysics Laboratory, NASA Goddard Space Flight Center, Code 662, Greenbelt, MD 20771, USA}

\author[0000-0003-2745-753X]{Paul T. Baker}
\affiliation{Department of Physics and Astronomy, Widener University, One University Place, Chester, PA 19013, USA}

\author[0000-0003-3053-6538]{Paul R. Brook}
\affiliation{Institute for Gravitational Wave Astronomy and School of Physics and Astronomy, University of Birmingham, Edgbaston, Birmingham B15 2TT, UK}

\author[0000-0002-6039-692X]{H. Thankful Cromartie}
\affiliation{National Research Council Research Associate, National Academy of Sciences, Washington, DC 20001, USA resident at Naval Research Laboratory, Washington, DC 20375, USA}

\author[0000-0002-1529-5169]{Kathryn Crowter}
\affiliation{Department of Physics and Astronomy, University of British Columbia, 6224 Agricultural Road, Vancouver, BC V6T 1Z1, Canada}

\author[0000-0002-2185-1790]{Megan E. DeCesar}
\altaffiliation{Resident at the Naval Research Laboratory}
\affiliation{Department of Physics and Astronomy, George Mason University, Fairfax, VA 22030, resident at the U.S. Naval Research Laboratory, Washington, DC 20375, USA}

\author[0000-0002-6664-965X]{Paul B. Demorest}
\affiliation{National Radio Astronomy Observatory, 1003 Lopezville Rd., Socorro, NM 87801, USA}

\author[0000-0001-8885-6388]{Timothy Dolch}
\affiliation{Department of Physics, Hillsdale College, 33 E. College Street, Hillsdale, MI 49242, USA}
\affiliation{Eureka Scientific, 2452 Delmer Street, Suite 100, Oakland, CA 94602-3017, USA}

\author[0000-0001-7828-7708]{Elizabeth C. Ferrara}
\affiliation{Department of Astronomy, University of Maryland, College Park, MD 20742, USA}
\affiliation{Center for Research and Exploration in Space Science and Technology, NASA/GSFC, Greenbelt, MD 20771}
\affiliation{NASA Goddard Space Flight Center, Greenbelt, MD 20771, USA}

\author[0000-0001-5645-5336]{William Fiore}
\affiliation{Department of Physics and Astronomy, University of British Columbia, 6224 Agricultural Road, Vancouver, BC V6T 1Z1, Canada}

\author[0000-0001-8384-5049]{Emmanuel Fonseca}
\affiliation{Department of Physics and Astronomy, West Virginia University, P.O. Box 6315, Morgantown, WV 26506, USA}
\affiliation{Center for Gravitational Waves and Cosmology, West Virginia University, Chestnut Ridge Research Building, Morgantown, WV 26505, USA}

\author[0000-0001-7624-4616]{Gabriel E. Freedman}
\affiliation{NASA Goddard Space Flight Center, Greenbelt, MD 20771, USA}

\author[0000-0001-6166-9646]{Nate Garver-Daniels}
\affiliation{Department of Physics and Astronomy, West Virginia University, P.O. Box 6315, Morgantown, WV 26506, USA}
\affiliation{Center for Gravitational Waves and Cosmology, West Virginia University, Chestnut Ridge Research Building, Morgantown, WV 26505, USA}

\author[0000-0001-8158-683X]{Peter A. Gentile}
\affiliation{Department of Physics and Astronomy, West Virginia University, P.O. Box 6315, Morgantown, WV 26506, USA}
\affiliation{Center for Gravitational Waves and Cosmology, West Virginia University, Chestnut Ridge Research Building, Morgantown, WV 26505, USA}

\author[0000-0003-1884-348X]{Deborah C. Good}
\affiliation{Department of Physics and Astronomy, University of Montana, 32 Campus Drive, Missoula, MT 59812}

\author[0000-0003-2742-3321]{Jeffrey S. Hazboun}
\affiliation{Department of Physics, Oregon State University, Corvallis, OR 97331, USA}

\author[0000-0001-6607-3710]{Megan L. Jones}
\affiliation{Center for Gravitation, Cosmology and Astrophysics, Department of Physics and Astronomy, University of Wisconsin-Milwaukee,\\ P.O. Box 413, Milwaukee, WI 53201, USA}

\author[0000-0001-6295-2881]{David L. Kaplan}
\affiliation{Center for Gravitation, Cosmology and Astrophysics, Department of Physics and Astronomy, University of Wisconsin-Milwaukee,\\ P.O. Box 413, Milwaukee, WI 53201, USA}

\author[0000-0002-0893-4073]{Matthew Kerr}
\affiliation{Space Science Division, Naval Research Laboratory, Washington, DC 20375-5352, USA}

\author[0000-0003-0721-651X]{Michael T. Lam}
\affiliation{SETI Institute, 339 N Bernardo Ave Suite 200, Mountain View, CA 94043, USA}
\affiliation{School of Physics and Astronomy, Rochester Institute of Technology, Rochester, NY 14623, USA}
\affiliation{Laboratory for Multiwavelength Astrophysics, Rochester Institute of Technology, Rochester, NY 14623, USA}

\author{T. Joseph W. Lazio}
\affiliation{Jet Propulsion Laboratory, California Institute of Technology, 4800 Oak Grove Drive, Pasadena, CA 91109, USA}

\author[0000-0003-1301-966X]{Duncan R. Lorimer}
\affiliation{Department of Physics and Astronomy, West Virginia University, P.O. Box 6315, Morgantown, WV 26506, USA}
\affiliation{Center for Gravitational Waves and Cosmology, West Virginia University, Chestnut Ridge Research Building, Morgantown, WV 26505, USA}

\author[0000-0001-5373-5914]{Jing Luo}
\altaffiliation{Deceased}
\affiliation{Department of Astronomy \& Astrophysics, University of Toronto, 50 Saint George Street, Toronto, ON M5S 3H4, Canada}

\author[0000-0001-5229-7430]{Ryan S. Lynch}
\affiliation{Green Bank Observatory, P.O. Box 2, Green Bank, WV 24944, USA}

\author[0000-0001-5481-7559]{Alexander McEwen}
\affiliation{Center for Gravitation, Cosmology and Astrophysics, Department of Physics and Astronomy, University of Wisconsin-Milwaukee,\\ P.O. Box 413, Milwaukee, WI 53201, USA}

\author[0000-0002-4642-1260]{Natasha McMann}
\affiliation{Department of Physics and Astronomy, Vanderbilt University, 2301 Vanderbilt Place, Nashville, TN 37235, USA}

\author[0000-0001-8845-1225]{Bradley W. Meyers}
\affiliation{Australian SKA Regional Centre (AusSRC), Curtin University, Bentley, WA 6102, Australia}
\affiliation{International Centre for Radio Astronomy Research (ICRAR), Curtin University, Bentley, WA 6102, Australia}

\author[0000-0002-3616-5160]{Cherry Ng}
\affiliation{Dunlap Institute for Astronomy and Astrophysics, University of Toronto, 50 St. George St., Toronto, ON M5S 3H4, Canada}

\author[0000-0002-6709-2566]{David J. Nice}
\affiliation{Department of Physics, Lafayette College, Easton, PA 18042, USA}

\author[0000-0001-5465-2889]{Timothy T. Pennucci}
\affiliation{Institute of Physics and Astronomy, E\"{o}tv\"{o}s Lor\'{a}nd University, P\'{a}zm\'{a}ny P. s. 1/A, 1117 Budapest, Hungary}

\author[0000-0002-8509-5947]{Benetge B. P. Perera}
\affiliation{Arecibo Observatory, HC3 Box 53995, Arecibo, PR 00612, USA}

\author[0000-0002-8826-1285]{Nihan S. Pol}
\affiliation{Department of Physics, Texas Tech University, Box 41051, Lubbock, TX 79409, USA}

\author[0000-0002-2074-4360]{Henri A. Radovan}
\affiliation{Department of Physics, University of Puerto Rico, Mayag\"{u}ez, PR 00681, USA}

\author[0000-0001-5799-9714]{Scott M. Ransom}
\affiliation{National Radio Astronomy Observatory, 520 Edgemont Road, Charlottesville, VA 22903, USA}

\author[0000-0002-5297-5278]{Paul S. Ray}
\affiliation{Space Science Division, Naval Research Laboratory, Washington, DC 20375-5352, USA}

\author[0000-0003-4391-936X]{Ann Schmiedekamp}
\affiliation{Department of Physics, Penn State Abington, Abington, PA 19001, USA}

\author[0000-0002-1283-2184]{Carl Schmiedekamp}
\affiliation{Department of Physics, Penn State Abington, Abington, PA 19001, USA}

\author[0000-0002-7283-1124]{Brent J. Shapiro-Albert}
\affiliation{Department of Physics and Astronomy, West Virginia University, P.O. Box 6315, Morgantown, WV 26506, USA}
\affiliation{Center for Gravitational Waves and Cosmology, West Virginia University, Chestnut Ridge Research Building, Morgantown, WV 26505, USA}
\affiliation{Giant Army, 915A 17th Ave, Seattle WA 98122}

\author[0000-0001-9784-8670]{Ingrid H. Stairs}
\affiliation{Department of Physics and Astronomy, University of British Columbia, 6224 Agricultural Road, Vancouver, BC V6T 1Z1, Canada}

\author[0000-0002-7261-594X]{Kevin Stovall}
\affiliation{National Radio Astronomy Observatory, 1003 Lopezville Rd., Socorro, NM 87801, USA}

\author[0000-0002-2820-0931]{Abhimanyu Susobhanan}
\affiliation{Max-Planck-Institut f{\"u}r Gravitationsphysik (Albert-Einstein-Institut), Callinstra{\ss}e 38, D-30167 Hannover, Germany\\}

\author[0000-0002-1075-3837]{Joseph K. Swiggum}
\altaffiliation{NANOGrav Physics Frontiers Center Postdoctoral Fellow}
\affiliation{Department of Physics, Lafayette College, Easton, PA 18042, USA}

\author[0000-0001-9678-0299]{Haley M. Wahl}
\affiliation{Department of Physics and Astronomy, West Virginia University, P.O. Box 6315, Morgantown, WV 26506, USA}
\affiliation{Center for Gravitational Waves and Cosmology, West Virginia University, Chestnut Ridge Research Building, Morgantown, WV 26505, USA}

\collaboration{1000}{(NANOGrav collaboration)}

\correspondingauthor{Lankeswar Dey}
\email{lankeswar.dey@nanograv.org}

\begin{abstract}

Free-floating objects (FFOs) in interstellar space---rogue planets, brown dwarfs, and large asteroids that are not gravitationally bound to any star---are expected to be ubiquitous throughout the Milky Way. 
Recent microlensing surveys have discovered several free-floating planets that are not bound to any known stellar systems. 
Additionally, three interstellar objects, namely 1I/'Oumuamua, 2I/Borisov, and 3I/ATLAS, have been detected passing through our solar system on hyperbolic trajectories. 
In this work, we search for FFOs on hyperbolic orbits that pass near millisecond pulsars (MSPs), where their gravitational influence can induce detectable perturbations in pulse arrival times. 
Using the NANOGrav 15-year narrowband dataset, which contains high-precision timing data for 68 MSPs, we conduct a search for such hyperbolic scattering events between FFOs and pulsars. 
Although no statistically significant events were detected, this non-detection enables us to place upper limits on the number density of FFOs as a function of their mass within our local region of the Galaxy.
For example, the upper limit on the number density for Jupiter-mass FFOs ($\sim 10^{-2.5} - 10^{-3.5}~M_{\odot}$) obtained from different pulsars ranges from $5.25\times10^{6}~\text{pc}^{-3}$ to $5.37\times10^{9}~\text{pc}^{-3}$, while the upper limit calculated by combining results from all the pulsars is $6.03\times10^{5}~\text{pc}^{-3}$.
These results represent the first constraints on FFO population derived from pulsar timing data.

\end{abstract}


\section{Introduction} \label{sec:intro}

Free-floating objects (FFOs) are non-stellar celestial bodies---including rogue planets, brown dwarfs, and large asteroid or cometary bodies---that reside in interstellar space and are not gravitationally bound to any stellar system. 
The first FFOs discovered were brown dwarfs and rogue planets, such as OTS~44 and Cha~110913--773444, which were ejected from their original stellar systems \citep{Luhman+2005_OTS44, Luhman+2005_Cha110913}.
Over the past decade, microlensing surveys have identified several free-floating planets (FFPs) with masses ranging from terrestrial to Jovian scales, as well as isolated brown dwarfs \citep{Sumi+2011_FFP, Mroz+2019_2new_FFP, Mroz+2020_teresrial_mass_FFP, Sumi+2023_FFP_mass_function, Gould+2009_BD_microlensing, Gould&Yee_2013_FFP_BD}. 
The first identified small extrasolar object passing through our solar system, i.e., interstellar object (ISO) was 1I/`Oumuamua, detected in 2017, followed by 2I/Borisov in 2019 \citep{Meech+2017_Oumuamua, Jewitt+2019_Borisov, Jewitt+2023_IS_interlopers}. 
Very recently, a third ISO 3I/ATLAS has been detected passing through the solar system in a hyperbolic orbit around the Sun with orbital eccentricity of $\sim 6.1$ \citep{Seligman+2025_ATLAS}.

FFOs are predicted to exist in large numbers across our Galaxy as models of planetary accretion and migration suggest that $\sim 75\%–85\%$ of cometary bodies  and $\sim40\%-80\%$ of the planets initially formed in a stellar system are scattered into interstellar space \citep{McGlynn+1989_ISOs, Brasser+2006_ISO_formation, Scholz+2022_FFP_BD_population}. 
Estimates place the local ISO number density near the Sun at approximately $10^{15} \, \text{pc}^{-3}$, and we expect ISOs to the be distributed across the entire Galaxy \citep{Engelhardt+2017_ISO_UL, Do+2018_ndensity_oumuamua, Forbes2025}. 
\cite{Coleman+2025_FFP_polulation_simulation} predicted based on simulations that on average there are 2.16 FFPs with masses $> 3\times10^{-8}\,\text{M}_{\odot}$ per star.
\cite{Sumi+2023_FFP_mass_function} found the number of FFPs per star to be $\sim 21$ over the mass range $10^{-6} < \text{M/M}_{\odot} < 0.02$ based on MOA-II microlensing survey in 2006$-$2014.
Millisecond pulsars (MSPs), rapidly rotating and highly magnetized neutron stars, also offer a promising avenue to probe the existence and population of FFOs in our Galaxy.

MSPs serve as highly precise celestial clocks due to their remarkably stable rotation periods. 
When observed with radio telescopes, the times of arrival (TOAs) of their pulses can be measured with exceptional ($\sim$ sub-microsecond) accuracy. 
These TOAs are used to construct a timing model, which accounts for every single rotation of the pulsar through a technique known as pulsar timing \citep{Lorimer_Kramer_hpa}. 
The timing model also enables the prediction of future TOAs, and the differences between the observed and predicted TOAs, known as timing residuals, can provide valuable insights. 
Pulsar timing has been instrumental in studying the properties of the interstellar medium, testing general relativity in the strong-field regime, and exploring low-frequency gravitational waves (GWs), among many other applications \citep{Lorimer_Kramer_hpa, Kramer+2021_double_psr_test_GR, IPTA_2024_3P+, NG12p5_2024_3C66B_ecc}.
Furthermore, pulsar timing can be employed to detect FFOs passing in close proximity to an MSP in our Galaxy \citep{Jennings_2020_ISO_Psr_scattering}.

\citet{Mitrofanov1990} proposed that the passage of an interstellar comet through the magnetosphere of a `dead' pulsar could potentially trigger a gamma-ray burst (GRB). Later, \citet{Brook+2014} suggested that an encounter between PSR~J0738$-$4042 and an asteroid might have caused the multiple pulse shape variations and the sudden change in the pulsar’s spin-down rate observed in 2005.
Gravitational scattering events, in which the path of an FFO is deflected by a pulsar, would slightly alter the motion of the pulsar relative to the solar system barycenter and therefore produce a detectable perturbation of the pulse arrival times.
\cite{Jennings_2020_ISO_Psr_scattering} calculated the shape of the expected timing perturbation induced by such encounters and explored the possibility of detecting them in pulsar timing array (PTA) datasets.
PTA experiments regularly observe an ensemble of MSPs in order to detect nanohertz (nHz) GWs \citep{Burke-Spolaor+2019}. 
Recently PTA collaborations across the world have found strong evidence for the presence of a nHz GW background in the universe \citep{NG15_2023_GWB, EPTA_DR2_2023_GWB, PPTA_DR3_2023_GWB, CPTA_2023_GWB, MPTA_2025_GWB}.
The long-term and high-precision pulsar timing data from these experiments provide an ideal resource to search for hyperbolic gravitational scattering events involving FFOs and pulsars in our Galaxy.

Primordial black holes (PBHs), hypothesized to have formed in the early universe shortly after the Big Bang, have been proposed as potential constituents of dark matter and may be distributed throughout our Galaxy \citep{Hawking_1971_PBH, Framptom+2010_PBH_DM, Carr&Kuhnel_2021_PBH}. 
Due to their lack of electromagnetic emission or reflection, PBHs are inherently difficult to detect via conventional observational techniques. 
Recently, \citet{Brown+2025_PBH_exoplanet} investigated the possibility that close encounters with PBHs could perturb exoplanetary orbits from their initial configurations. 
Similarly, because pulse arrival time delays induced by the hyperbolic scattering of an object near a pulsar depend purely on gravitational interactions, this method offers a promising avenue for identifying PBHs in the Milky Way and probing their population properties. 
Although PBHs are generally not classified as part of the FFO population, in this work, we consider any celestial object not gravitationally bound to a star and traveling through interstellar space as an FFO.

In this paper, we use the North American Nanohertz Observatory for Gravitational Waves (NANOGrav) 15-year dataset \citep{NG2023_15yr_timing} to search for gravitational scattering of FFOs by pulsars.
This dataset contains timing data for 68 MSPs with an observation span of up to 15.9 years.

The paper is organized as follows. 
In Section~\ref{sec:signal_model}, we describe the model used to calculate the pulse arrival time perturbations induced by the hyperbolic gravitational scattering of an FFO by a pulsar. 
Section~\ref{sec:data_methods} provides a brief overview of the data utilized, the complete model employed to fit the pulsar TOAs, and the priors assigned to the fitting parameters. 
The results of our study are presented in Section~\ref{sec:results}, and a summary of our findings is provided in Section~\ref{sec:summary}.

Throughout the paper, the logarithm function ($\log$) has been used with a base of 10 unless explicitly stated otherwise.

\section{Signal Model}
\label{sec:signal_model}

When an FFO or similar perturbing object with mass $m$ passes close to a pulsar on a hyperbolic orbit with impact parameter $b$ and eccentricity $e$, the excess time delay induced in pulse arrival time due to the reflex motion of the pulsar at a time $t$ can be written as (see \citealt{Jennings_2020_ISO_Psr_scattering} for a detailed derivation)
\begin{align}
    R(t) = \frac{m \sin{i}\ b}{Mc} \left( \frac{e - \cosh{H(t)}}{\sqrt{e^2 - 1}}  \sin{\omega} + \sinh{H(t)} \cos{\omega} \right)\,,
    \label{eqn:residual}
\end{align}
where $M$ is the total mass of the pulsar and the perturber, $i$ is the inclination angle of the scattering plane with respect to the sky plane (not to be confused with orbital plane in case of binary pulsars), $\omega$ is the argument of periapsis, $H(t)$ is the hyperbolic anomaly, and $c$ is the speed of light in vacuum.
The geometry of these parameters is shown in Figure~1 of \cite{Jennings_2020_ISO_Psr_scattering}.
As all the perturbing objects considered in this paper have masses significantly less than $1\,M_{\odot}$, and the exact masses of most of the pulsars are not precisely known, we approximated the total mass $M$ to be equal to the approximate mass of the pulsar ($\sim1.4\,M_{\odot}$) for isolated pulsars \citep{Thorsett&Chakrabarty_1999_pulsar_mass}. 
For pulsars with a companion, we use the total mass of the pulsar binary system as $M$ (see Section~\ref{subsubsec:prior_binary} for details).

The eccentricity $e$ is related to the asymptotic velocity $v_{\infty}$ of the perturber relative to the pulsar by:
\begin{align}
    e = \sqrt{1 + \left(\frac{b\,v_{\infty}^2}{GM}\right)^2}\,,
    \label{eqn:e_b_relation}
\end{align}
where $G$ is the universal gravitational constant.
The hyperbolic anomaly $H$ is related to the time $t$ by the hyperbolic Kepler equation:
\begin{align}
    t = t_0 + \frac{b\,(e\,\sinh{H(t)} - H(t))}{v_{\infty}\,\sqrt{e^2 - 1}}\,,
    \label{eqn:hyp_KE}
\end{align}
where $t_0$ is the time of periapsis (point of closest approach) at which $H = 0$.
As long as $e>1$, as it is for a hyperbolic orbit, $t$ is an increasing function of $H$, and Equation~(\ref{eqn:hyp_KE}) can be inverted to get $H(t)$ as a function of $t$.

Further, using the conservation of energy and angular momentum, we can represent $b$ in terms of the periapsis distance $p$ and asymptotic velocity $v_{\infty}$ as:
\begin{align}
    b = p \, \sqrt{1 + \frac{2GM}{p\,v^2_{\infty}}}\,.
    \label{eqn:b_p_relation}
\end{align}
We use the parameters $p$ and $v_{\infty}$ as our independent model parameters and calculate $b$ and $e$ from them using Equations~\eqref{eqn:b_p_relation} and \eqref{eqn:e_b_relation} while calculating $R(t)$.
This choice of independent parameters is motivated by the priors we need to set for the parameter space in our search and is discussed further in Section~\ref{subsec:priors}.
Therefore, if the pulsar mass is known, the induced arrival time perturbations can be precisely calculated given the values for the parameters $m$, $i$, $p$, $v_{\infty}$, $\omega$, and $t_0$.

\begin{figure*}
    \centering
    \includegraphics[width=0.48\linewidth]{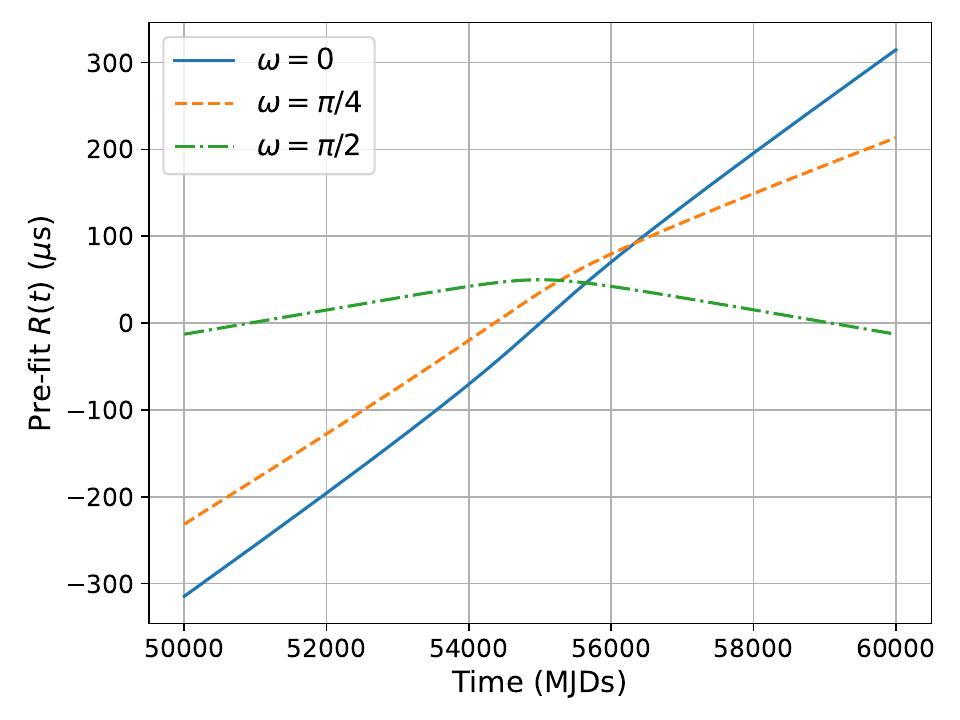}
    \includegraphics[width=0.48\linewidth]{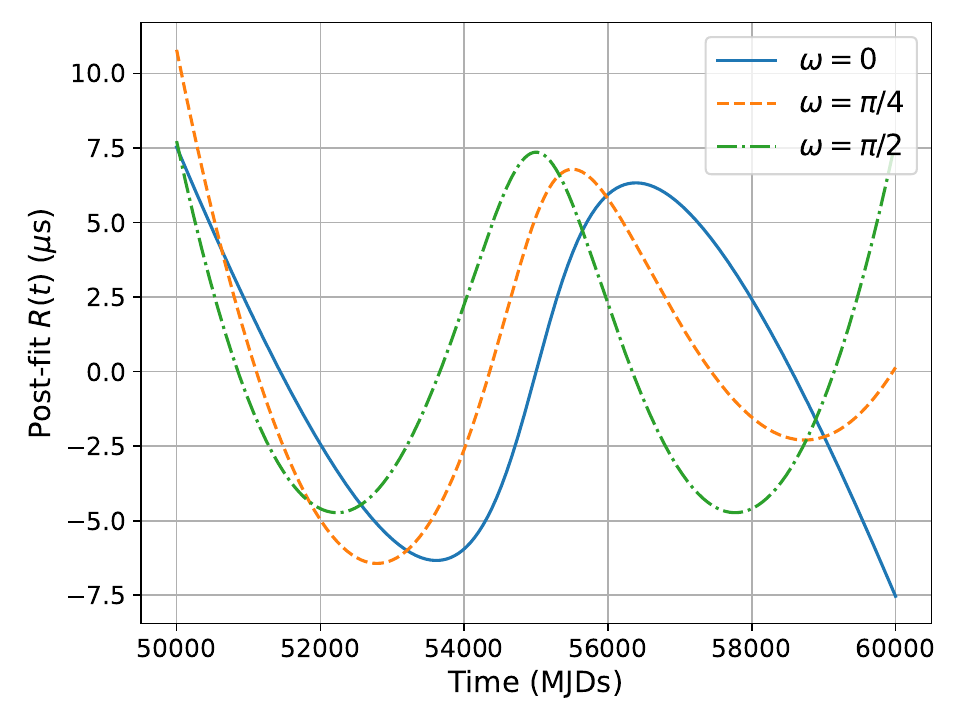}
    \caption{Pulse arrival time perturbations induced by a hyperbolic gravitational scattering of an FFO with projected mass $m\sin{i} = 10^{-8}\,M_{\odot}$, periapsis distance $p = 10$ AU, asymptotic velocity relative to the pulsar $v_{\infty} = 20$ km/s, and time of periapsis $t_0 = 55000$ MJD, for a pulsar with mass $M = 1.4 \,M_{\odot}$.
    The left panel displays the total induced time delay due to the reflex motion of the pulsar and the right panel shows these time perturbations after subtracting a second-order polynomial fit, for three different orientations of the orbit.}
    \label{fig:residuals}
\end{figure*}

Since the mass $m$ of the perturber and the inclination angle $i$ of the hyperbolic orbit appear together as a multiplicative factor in Equation~(\ref{eqn:residual}) and do not influence other parts of the equation independently, we treat the projected perturber mass $m\sin{i}$ as a single parameter instead of separating $m$ and $i$ during our search for such signals in the data. 
Additionally, the excess time delay, $R(t)$, induced by a hyperbolic scattering event may include significant linear and quadratic components as functions of time, which are degenerate with the contributions of the pulsar's spin period and spin period time derivative \citep[see Section 4 of][]{Jennings_2020_ISO_Psr_scattering}. 
To mitigate this degeneracy, we fit and remove a second-order polynomial from the time-delay models R(t) used in this search.
Figure~\ref{fig:residuals} illustrates an example of pulse arrival time delays caused by the gravitational scattering of an FFO, both before and after subtracting the second-order polynomial fit.

Many of the MSPs analyzed in this study to search for FFOs are part of binary systems, with typical orbital separations ranging from $0.01$ to $0.6$ AU. 
A potential complication arises if an FFO passes within a few times the binary separation, as three-body interactions could significantly influence the system's dynamics and the resulting pulse travel times. 
Accurately modeling such three-body interactions is complex and beyond the scope of this paper. 
Consequently, our analysis is restricted to a parameter space where the effects of three-body interactions are negligible. 
A detailed discussion of these considerations and the corresponding parameter space selection is provided in Section~\ref{subsec:priors}.

\section{Data and Analysis Methods}
\label{sec:data_methods}

In this section, we provide a brief overview of the pulsar timing dataset utilized in this study, followed by a detailed description of the complete model used to fit the times of arrival of pulses from a pulsar. 
Additionally, we outline the priors for the various free model parameters and describe the detection statistics employed in our analysis.

\subsection{NANOGrav 15-year Dataset}
\label{subsec:dataset}

We used the NANOGrav 15-year narrowband dataset \citep{NG2023_15yr_timing} to search for gravitational scattering of FFOs by pulsars in our Galaxy.
This dataset includes sub-banded TOAs and best-fit timing models for 68 MSPs, with a maximum data span of 15.9 years for some sources. 
Observations were conducted using the Arecibo Observatory, the Green Bank Telescope, and the Very Large Array at frequencies ranging from 327 MHz to 3 GHz, with most pulsars observed roughly once per month. 
All timing analyses employed the JPL DE440 Solar System ephemeris and the TT(BIPM2019) timescale.
Dispersion measure (DM) variations in the TOAs were corrected using the DMX model, which provides a piecewise constant representation of DM changes. 
Among the 68 MSPs, 50 have known binary companions. 
For this study, we individually analyzed the timing data for all 68 pulsars to search for hyperbolic gravitational scattering events involving an FFO and the pulsar within the data span.

\subsection{The Search Process}

Pulsar timing residuals denote the deviation of the observed TOAs from those predicted by a timing model.
While searching for a gravitational scattering events by a pulsar, we begin with an original solution as published in \cite{NG2023_15yr_timing}.
We model perturbations to that solution due to scattering by an FFO and also deviations in the original timing model due to covariances between the FFO perturbations and other timing phenomena, like pulsar spin frequency and its time derivative. 
The difference between the timing residuals of the original solution and the timing residuals of the new solution including the FFO-induced delay is then:
\begin{align}
    r(t) = \text{TM}(t) + \text{WN}(t) + \text{RN}(t) + R(t) \,,
\end{align}
where TM represents a marginalized timing model that can account for deviations of the timing model parameters from their original values, WN and RN stand for white noise and red noise for the pulsar, respectively, and $R(t)$ denotes the timing residual induced by the scattering event, as defined in Equation~\eqref{eqn:residual}.

The  WN is characterized using three phenomenological parameters: EFAC, EQUAD, and ECORR, one of each defined for every telescope receiver-backend combination. 
The EFAC parameter scales the TOA measurement uncertainties by a multiplicative factor, while the EQUAD parameter adds in quadrature to the measurement uncertainties. 
The ECORR parameter accounts for radio frequency-correlated WN between narrowband TOAs derived from the same observation \citep[see][]{NG2023_15yr_noise}.
In our search, we keep the WN parameters fixed to the maximum-likelihood values obtained from individual pulsar noise analysis and do not treat them as free parameters.

The RN for each pulsar is modeled as a Gaussian red noise process with a power-law spectral density
\begin{align}
    P_{\text{RN}} = \frac{A^2_{\text{RN}}}{12\pi^2} \left(\frac{f}{f_{\rm yr}}\right)^{-\gamma_{\text{RN}}} \text{yr}^3 \,, 
\end{align}
where $A_{\text{RN}}$ and $\gamma_{\text{RN}}$ represent the RN amplitude and spectral index, respectively, and $f_{\rm yr} = \text{yr}^{-1}$.
Following \cite{NG2023_15yr_timing} and \cite{NG2023_15yr_noise}, we model the RN of each pulsar as a Fourier sum comprising 30 frequency components ranging from $1/T_{\text{span}}$ to $30/T_{\text{span}}$, where $T_{\text{span}}$ is the observation span of the pulsar.

The timing residual signal $R(t)$, induced by hyperbolic gravitational scattering events---such as those illustrated in Figure~\ref{fig:residuals}---can closely resemble a realization of the red-noise process, $\text{RN}(t)$. 
To examine whether $R(t)$ can be fully absorbed by $\text{RN}(t)$ when both components are modeled simultaneously in a pulsar timing analysis, we performed a series of simulation studies. 
To generate the simulated datasets, we included only white noise and a gravitational scattering signal $R(t)$, added to an idealized set of pulsar TOAs. 
To ensure detectability of the scattering signal, we set its amplitude to approximately $R(t) \sim 1~\mu\text{s}$. We then searched for the combined model $R(t) + \text{RN}(t)$ in the simulated datasets while fixing the white noise parameters to their injected values.

Across all simulations, we successfully recovered the injected gravitational scattering signal, while the Bayes factor for red-noise detection remained below unity. 
This indicates that, in the absence of an intrinsic red-noise process in the pulsar TOAs, our search framework does not favor the red-noise model over the gravitational-scattering model when both the deterministic scattering signal and the stochastic red-noise component are included in the analysis.

In summary, our search model consists of seven parameters that are sampled using a Markov Chain Monte Carlo approach to fit the pulsar timing residuals within a Bayesian framework, while marginalizing over the free parameters of the timing model.
These seven parameters include two RN parameters--- $A_{\text{RN}}$ and $\gamma_{\text{RN}}$, and five more parameters--- $m\sin{i}$, $p$, $v_{\infty}$, $\omega$, and $t_0$---which are used to model $R(t)$. 
The priors assigned to each of these parameters are discussed in the subsequent section.

\subsection{Priors}
\label{subsec:priors}

As mentioned earlier, for the white noise parameters EFAC, EQUAD, and ECORR, we fix their values to the maximum-likelihood estimates obtained from the single-pulsar noise analysis in \cite{NG2023_15yr_timing}. 
This approach is standard in PTA analyses, and fixing these parameters is not expected to affect our results compared to a scenario where they are treated as free parameters in the search for an FFO-scattering signal in our data \citep{NG15_2023_GWB}.
For the red noise parameters, we adopt a log-uniform prior for $A_{\text{RN}}$ and a uniform prior for $\gamma_{\text{RN}}$ with the following bounds: $\log{A_{\text{RN}}} \in U[-18, -11]$ and $\gamma_{\text{RN}} \in U[0, 7]$.

The priors used for the parameters determining the hyperbolic orbit, along with those for the red noise parameters, are summarized in Table~\ref{tab:priors}. 
For the projected mass of the FFO, we adopt a log-uniform prior with a lower bound of $10^{-14}\,M_{\odot}$ and an upper bound of $0.07\,M_{\odot}$. 
This lower bound is chosen because an FFO with a projected mass below this value would not produce a detectable signal in the pulsar timing given our current sensitivity and the range of $p$ considered in this study. 
The upper bound corresponds to the upper limit mass definition of a brown-dwarf $\approx 0.07 M_{\odot}$ \citep{Chabrier_2023_HBL}.
For now, we are ignoring the $\sin{i}$ factor while choosing the prior, but the effect of this factor in our results is discussed later in Section~\ref{subsec:FFO_density}.

The prior for $v_{\infty}$, the asymptotic velocity of the FFO relative to the pulsar, is chosen to follow a Maxwellian distribution, with the probability density function given by
\begin{align}
p(v_{\infty})\,dv_{\infty} = \sqrt{\frac{2}{\pi}} \frac{v_{\infty}^2}{a^3} \exp\left(-\frac{v_{\infty}^2}{2a^2}\right)\,dv_{\infty},
\end{align}
where $a$ is the scale parameter of the distribution.
This choice is motivated by the fact that the velocity distributions of MSPs in the Milky Way are expected to be approximately Maxwellian \citep{Hobbs+2005}.
In this work, we adopt a scale parameter of $a = 55$ km/s, which corresponds to a mean asymptotic velocity $v_{\infty}$ to be equal to the mean 2D velocity of recycled MSPs ($\approx 87$ km/s) found by \cite{Hobbs+2005}.
Recent studies also found the mean of the velocity distribution of MSPs to have similar values \citep{Matthews+2016_NG9_astrometry,Shamohammadi+2024_MPTA_astrometry}.
The FFO velocities are expected to be sufficiently smaller than MSP velocities \citep{Hurley&Shara_2002_FFP_velocity, Coleman_2024_FFP_velocities, Forbes2025} and therefore can be safely ignored while choosing the prior for $v_{\infty}$.
Note that, the prior distribution of $v_{\infty}$ used in this paper is just an approximation to the expected velocity distribution of FFOs relative to MSPs in our galaxy, and small changes in the scale parameter $a$ should not substantially affect our results.

\begin{deluxetable*}{l l c c}
\tablecaption{Prior distribution of the model parameters \label{tab:priors}}
\tablewidth{0.8\textwidth}
\tablehead{
    \colhead{Parameter} &
    \colhead{Description} &
    \colhead{Unit} &
    \colhead{Prior}
    }
\startdata
$\log{A_{\text{RN}}}$ & Amplitude of the pulsar red noise process &  & Uniform[--18, --11]\\[2pt]
$\gamma_{\text{RN}}$ & Spectral index of the pulsar red noise process &  & Uniform[0, 7]\\[5pt]
$\log{(m\sin{i})}$ & Projected mass of the FFO & $M_{\odot}$ & Uniform[--14, --1.155]\\[2pt]
$\omega$ & Argument of periapsis & rad  & Uniform[0, $2\pi$]\\[2pt]
$v_{\infty}$ & Asymptotic velocity of the FFO & km/s & Maxwellian($a=55$ km/s)\\[2pt]
$t_0$ & Time of periapsis & MJD & Uniform[TOA$_{\text{min}}$, TOA$_{\text{max}}$]\tablenotemark{a}\\[5pt]
$\log{p}$ & Distance of periapsis of the hyperbolic orbit & AU & Uniform[--2.52, 6] for isolated pulsars\\
 &  &  & Uniform[$\log{p_{\text{min}}}$, 6] for pulsars in binary \tablenotemark{b}
\enddata
\tablenotetext{a}{TOA$_{\text{min}}$ and TOA$_{\text{max}}$ represent the earliest and latest TOAs for the pulsar, respectively.}
\tablenotetext{b}{See the last column of Table~\ref{tab:binaries} for the value of $\log{p_{\text{min}}}$ for each pulsar with a binary companion.}
\end{deluxetable*}

For the periapsis distance $p$, the prior distribution depends on whether the pulsar is isolated or part of a binary system.
For both isolated and binary pulsars, a log-uniform prior with an upper bound of $10^{6}$ AU is assigned to $p$. 
This upper bound is chosen because for $p > 10^{6}$ AU, the induced time delay becomes too small to be detectable for the mass range of FFOs considered in this study.
For isolated pulsars, a lower bound of 0.003 AU is set for the prior of $p$. 
Periapsis distances below 0.003 AU are excluded because tidal deformation becomes significant for a typical planet-like FFO at such close distances \citep{Rappaport2013}. 
For pulsars with binary companions, additional caution is required when selecting the lower limit of the prior for $p$ to ensure that the effects of three-body interactions remain negligible, as previously mentioned.
The detailed procedure for determining the lower limit on the prior for $p$ on a case-by-case basis for each pulsar with a binary companion is provided in the following sub-section.

\subsubsection{Periapsis Distance Priors for Binary Pulsars}
\label{subsubsec:prior_binary}

In the case of binary pulsars, we assume that for a sufficiently large $p$, the FFO does not interact with the pulsar and its companion separately but rather as one combined object placed at the pulsar-companion center of mass.
In other words, we treat the time delays due to the motion of this pulsar as the sum of two two-body interactions: the interaction with its companion and the interaction with the FFO. 
Therefore, 
\begin{equation}
    R_{\text{tot}}(t) = R_{\text{c}}(t) + R_{\text{FFO}}(t) + R_3(t),
\label{eq:3bodymodel}
\end{equation}
where $R_{\text{c}}(t)$ encapsulates the pulsar-companion closed orbit and $R_{\text{FFO}}(t)$ models the hyperbolic orbit and is equal to Equation~\eqref{eqn:residual} where $M$ is now the sum of the pulsar mass, M$_{\rm p}$, and companion mass, M$_2$. 
$R_3(t)$, which we call ``three-body interactions", describes any deviation from our double two-body model. 
$R_3(t)$ famously must be determined numerically through three-body simulations, as described in the following paragraphs. We define our $p$ lower-bound prior, $p_{\text{min}}$, as the $p$ such that 
\begin{equation}
    \mathcal{R}_3(p) \equiv \frac{\text{RMS}[R_{3}(t; p)]}{\text{RMS}[R_{\text{FFO}}(t; p)]} = 0.1,
\label{eq:normalized3body}
\end{equation}
where $\text{RMS}[f(t; p)]$ denotes the root-mean-square of the set $\{f(t_i) \text{ for all simulation time samples $t_i$}\}$ when the periapsis distance is set as $p$. 
$\mathcal{R}_3(p)$ is the normalized three-body interaction strength for a given $p$. 
We choose this criterion for $p_\text{min}$ to ensure that our search is restricted to periapsis distance values for which Equation~\eqref{eqn:residual} provides a reasonably accurate description of the perturbation in pulse arrival times resulting from an FFO scattering event when the pulsar has a binary companion.
The threshold value of 0.1 chosen for the ratio in Equation~\eqref{eq:normalized3body} is somewhat arbitrary, however, this choice does not influence the key results presented in this work.

We set up a three-body simulation for each binary pulsar in the NANOGrav 15-year dataset using the IAS15 integrator \citep{reboundias15} as implemented in the \texttt{rebound} Python package \citep{rebound}. 
We first initialize a closed binary system using the parameters listed in Table~\ref{tab:binaries} (see Table~\ref{tab:binaries} caption). 
If the pulsar mass is unknown or poorly constrained, we adopt a mass of 1.4~M$_\odot$.
We set the semi-major axis of the pulsar's orbit around the two-body center of mass as $A_1/ \sin{I}$, choosing $\sin{I} = 1$ when the inclination of the orbit is unknown for simplicity. 
Knowing or assigning $M_{\rm p}$, $\sin{I}$, and the binary orbital period ($P_{\text{b}}$) uniquely determines the companion mass via the binary mass function. 
We also set the closed-orbit eccentricity, $e_\text{closed}$, to 0 for simplicity.

We then add a third body---the FFO---on a hyperbolic trajectory to the simulation. 
We set the mutual inclination between the FFO and companion orbits to 0, $v_\infty=0.001$~km/s, $\omega=0$, $m~=~0.07~M_\odot$, which is the corner of the prior space that maximizes three-body interactions. 
We choose a simulation time span of 15~yr and $t_0 = 7.5$~yr. 
We then integrate the simulation with uniform time sampling for a range of $p$-values chosen so that $\mathcal{R}_3(p)$ crosses $\mathcal{R}_3=0.1$. 
The pulsar's motion along the line of sight such that $\omega=0$ represents $R_\text{tot}(t)$ in Equation \ref{eq:3bodymodel}. 
We fit $R_\text{tot}(t)$ for only $R_{\text{c}}(t) + R_{\text{FFO}}(t)$, and set $R_3(t) = R_\text{tot}(t) - [R_{\text{c, fit}}(t) + R_{\text{FFO, fit}}(t)]$. 
To allow for any induced eccentricity in the closed-orbit system, $R_{\text{c}}(t)$ is modeled by the ELL1 model as formulated in \citet{lange2001precision} which accounts for $\mathcal{O}(e_\text{closed})$ effects. 
We then follow Equation~\eqref{eq:normalized3body} to determine $\mathcal{R}_3(p)$. 
To find $p_\text{min}$ such that $\mathcal{R}_3(p_\text{min}) =0.1$, we interpolate by fitting an exponential profile to $\mathcal{R}_3(p)$. 
We do these sets of simulations for each binary pulsar in the NANOGrav 15-year dataset to set their $p_\text{min}$. 
The $p_\text{min}$ for each binary pulsar is listed in the last column of Table \ref{tab:binaries}.

\startlongtable
\begin{deluxetable*}{l c c c c c c}
\tablecaption{Here, we list selected orbital parameters of all binary pulsars included in the NANOGrav 15-year dataset. 
The second column represents the orbital period ($P_{\text{b}}$) of the binary while projected semi-major axis ($A_1 = A~\sin{I}$; $A$ is the semi-major axis) of the binary orbit is given in in third column.
The sine of the orbital inclination ($\sin{I}$) and the companion mass ($M_2$) are provided, where available, in the fourth and fifth columns, respectively. 
Additionally, we provide the mass of the pulsars for which we have well-constrained values derived from the timing of the NANOGrav 15-year data in the fifth column \citep[see Table 4 of][]{NG2023_15yr_timing}. 
Finally, the lower limit of the prior for $\log{p}$ for each pulsar is given in the last column. 
\label{tab:binaries}}
\tablehead{
    \colhead{PSR} &
    \colhead{$P_{\rm b}$ (days)} &
    \colhead{$A_1$ (ls)} &
    \colhead{$\sin{I}$} &
    \colhead{$M_2 (M_{\odot})$} & 
    \colhead{$M_{\rm p}$ ($M_{\odot}$)} &
    \colhead{$\log{p}_{\text{min}}$ (AU)}
    }
\startdata
B1855+09 & 12.327 & 9.231 & 0.999 & 0.262 & 1.563 & $-0.205$ \\
B1953+29 & 117.349 & 31.413 & $-$ & $-$ & $-$ & 0.223 \\
J0023+0923 & 0.139 & 0.035 & $-$ & $-$ & $-$ & $-1.499$ \\
J0406+3039 & 6.956 & 2.319 & $-$ & $-$ & $-$ & $-0.469$ \\
J0437$-$4715 & 5.741 & 3.367 & $-$ & $-$ & $-$ & $-0.447$ \\
J0509+0856 & 4.908 & 2.458 & $-$ & $-$ & $-$ & $-0.519$ \\
J0557+1551 & 4.847 & 4.054 & $-$ & $-$ & $-$ & $-0.470$ \\
J0605+3757 & 55.672 & 18.949 & $-$ & $-$ & $-$ & 0.062 \\
J0610$-$2100 & 0.286 & 0.073 & $-$ & $-$ & $-$ & $-1.314$ \\
J0613$-$0200 & 1.199 & 1.091 & 0.898 & 0.240 & - & $-0.944$ \\
J0614$-$3329 & 53.585 & 27.639 & $-$ & $-$ & $-$ & 0.098 \\
J0636+5128 & 0.067 & 0.009 & $-$ & $-$ & $-$ & $-1.741$ \\
J0709+0458 & 4.367 & 15.717 & $-$ & $-$ & $-$ & $-0.322$ \\
J0740+6620 & 4.767 & 3.978 & 0.999 & 0.247 & 1.99 & $-0.401$ \\
J1012+5307 & 0.605 & 0.582 & $-$ & $-$ & $-$ & $-0.980$ \\
J1012$-$4235 & 37.972 & 21.263 & $-$ & $-$ & $-$ & 0.022 \\
J1022+1001 & 7.805 & 16.765 & $-$ & $-$ & $-$ & $-0.274$ \\
J1125+7819 & 15.355 & 12.192 & $-$ & $-$ & $-$ & $-0.163$ \\
J1312+0051 & 38.504 & 14.750 & $-$ & $-$ & $-$ & $-0.019$ \\
J1455$-$3330 & 76.175 & 32.362 & $-$ & $-$ & $-$ & 0.164 \\
J1600$-$3053 & 14.348 & 8.802 & 0.912 & 0.235 & $-$ & $-0.242$ \\
J1614$-$2230 & 8.687 & 11.291 & 1.000 & 0.494 & 1.937 & $-0.206$ \\
J1630+3734 & 12.525 & 9.039 & 0.960 & 0.648 & $-$ & $-0.506$ \\
J1640+2224 & 175.461 & 55.330 & 0.874 & 0.453 & $-$ & 0.141 \\
J1643$-$1224 & 147.017 & 25.073 & $-$ & $-$ & $-$ & 0.418 \\
J1705$-$1903 & 0.184 & 0.104 & $-$ & $-$ & $-$ & $-1.339$ \\
J1713+0747 & 67.825 & 32.342 & $-$ & $-$ & $-$ & 0.116 \\
J1719$-$1438 & 0.091 & 0.002 & $-$ & $-$ & $-$ & $-1.860$ \\
J1738+0333 & 0.355 & 0.343 & $-$ & $-$ & $-$ & $-1.108$ \\
J1741+1351 & 16.335 & 11.003 & 0.964 & 0.202 & 0.83 & $-0.258$ \\
J1745+1017 & 0.730 & 0.088 & $-$ & $-$ & $-$ & $-1.148$ \\
J1751$-$2857 & 110.746 & 32.528 & $-$ & $-$ & $-$ & 0.224 \\
J1802$-$2124 & 0.699 & 3.719 & $-$ & $-$ & $-$ & $-0.797$ \\
J1811$-$2405 & 6.272 & 5.706 & 0.960 & 0.327 & $-$ & $-0.497$ \\
J1853+1303 & 115.654 & 40.770 & 0.873 & 0.170 & $-$ & 0.362 \\
J1903+0327 & 95.174 & 105.593 & 0.983 & 0.870 & 0.94 & 0.234 \\
J1909$-$3744 & 1.533 & 1.898 & 0.998 & 0.209 & 1.57 & $-0.699$ \\
J1910+1256 & 58.467 & 21.129 & $-$ & $-$ & $-$ & 0.082 \\
J1918$-$0642 & 10.913 & 8.350 & 0.996 & 0.225 & 1.31 & $-0.242$ \\
J1946+3417 & 27.020 & 13.869 & 0.922 & 0.672 & $-$ & $-0.451$ \\
J2017+0603 & 2.198 & 2.193 & 0.927 & 0.247 & $-$ & $-0.702$ \\
J2033+1734 & 56.308 & 20.163 & $-$ & $-$ & $-$ & 0.073 \\
J2043+1711 & 1.482 & 1.624 & 0.990 & 0.190 & 1.62 & $-0.862$ \\
J2145$-$0750 & 6.839 & 10.164 & $-$ & $-$ & $-$ & $-0.292$ \\
J2214+3000 & 0.417 & 0.059 & $-$ & $-$ & $-$ & $-1.290$ \\
J2229+2643 & 93.016 & 18.913 & $-$ & $-$ & $-$ & 0.139 \\
J2234+0611 & 32.001 & 13.937 & $-$ & $-$ & $-$ & $-0.054$ \\
J2234+0944 & 0.420 & 0.068 & $-$ & $-$ & $-$ & $-1.251$ \\
J2302+4442 & 125.935 & 51.430 & 0.994 & 0.262 & - & 0.317 \\
J2317+1439 & 2.459 & 2.314 & $-$ & $-$ & $-$ & $-0.592$ \\
\enddata
\end{deluxetable*}

\subsection{Detection and Upper Limit Statistics}

To evaluate the significance of detecting a hyperbolic gravitational scattering signal by an FFO for a pulsar, we compare the model $\mathcal{H}_1$ = TM + WN + RN + $R(t)$ against the model $\mathcal{H}_0$ = TM + WN + RN by computing the Bayes factor using the Savage-Dickey (SD) formula \citep{Dickey1971}.
The Savage-Dickey formula is applicable in this case because $\mathcal{H}_1$ and $\mathcal{H}_0$ are nested models, with $\mathcal{H}_0$ being the special case of $\mathcal{H}_1$ when $m\sin{i}=0$ in Equation~\eqref{eqn:residual} for $R(t)$.
The SD Bayes factor is expressed in this case by
\begin{align}
    \mathcal{B} = \frac{p[m\sin{i}=0|\mathcal{H}_1]}{p[m\sin{i}=0|\mathcal{D},\mathcal{H}_1]}\,,
\end{align}
where $\mathcal{D}$ represents the data, and $p[m\sin{i}=0|\mathcal{H}_1]$ and $p[m\sin{i}=0|\mathcal{D},\mathcal{H}_1]$ are the values of the marginalized prior distribution and the posterior distribution at $m\sin{i}=0$, respectively.
The uncertainty in this estimate is given by $\mathcal{B}/\sqrt{N_0}$ where $N_0$ denotes the number of samples in the lowest $m\sin{i}$ bin.

\begin{figure*}
    \centering
    \includegraphics[width=0.75\linewidth]{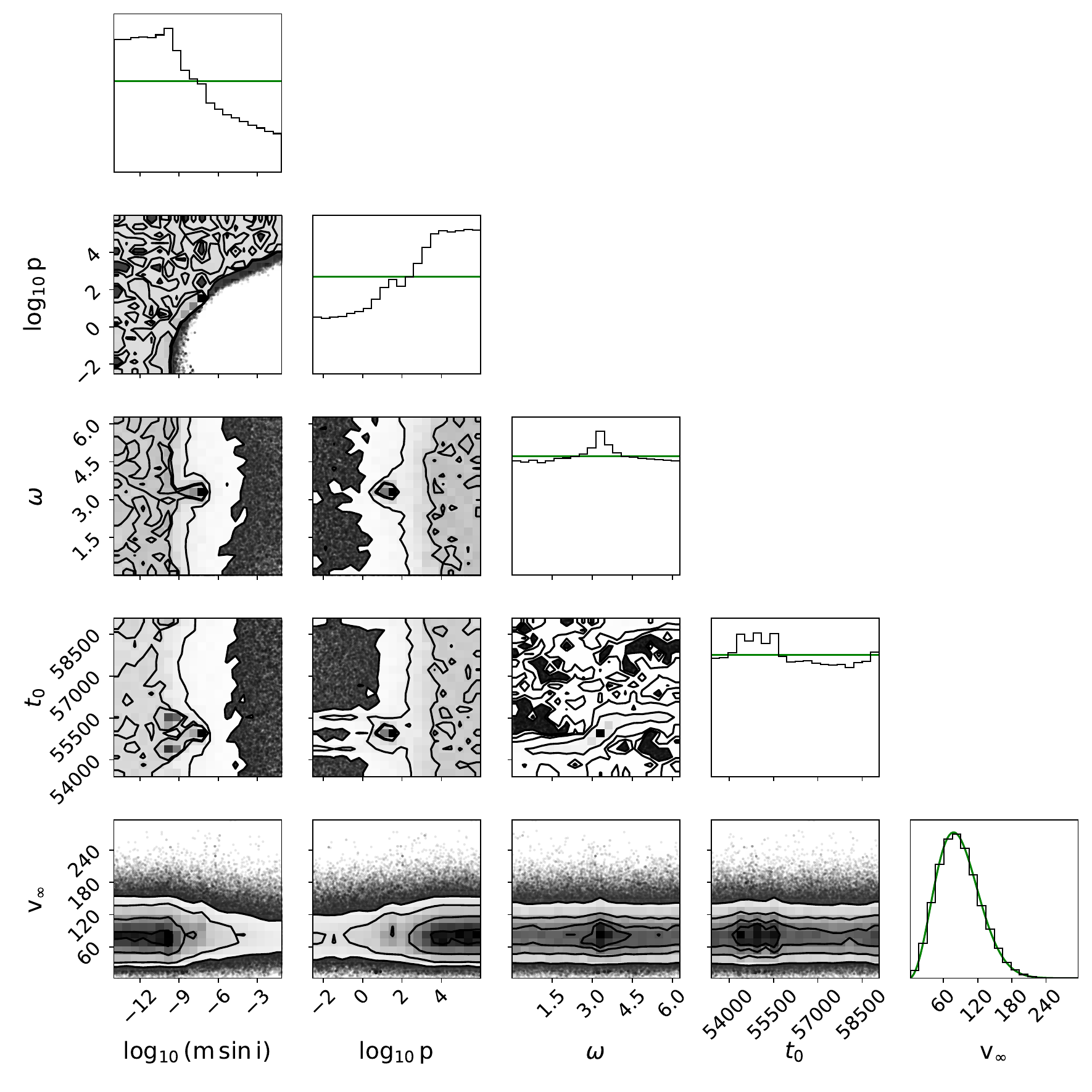}\\
    \includegraphics[width=0.75\linewidth]{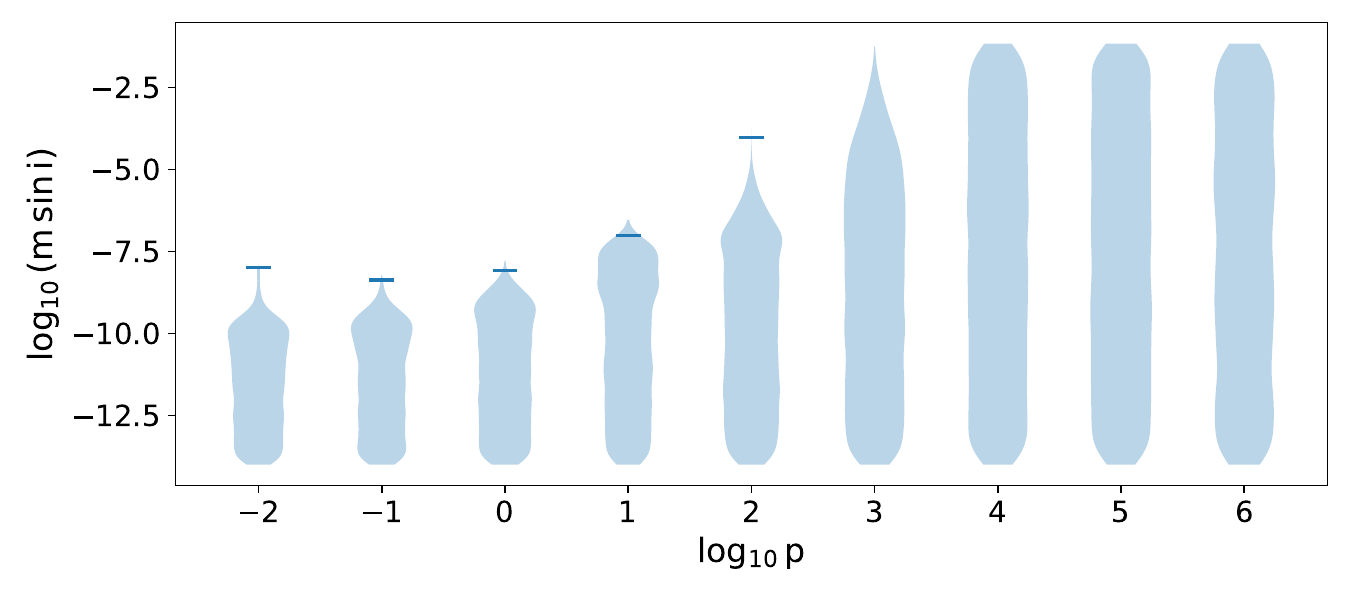}
    \caption{Top panel: Corner plot showing the 1D (black histograms) and 2D posterior distributions of the hyperbolic scattering model parameters for PSR J0030$+$0451.
    The priors assigned for each parameter are also shown by green lines in the 1D distributions.
    Bottom panel: Violin plot showing the binned posterior distributions of the FFO projected mass as a function of $\log{p}$ with 95\% ULs shown by the horizontal blue dashes for PSR J0030$+$0451.
    The units for $m\sin{i}$, $p$, $\omega$, $t_0$, and $v_{\infty}$ are $M_{\odot}$, AU, radian, MJD, and km/s, respectively.}
    \label{fig:posterior_ULs}
\end{figure*}

In case of a non-detection, i.e., for $\mathcal{B} \lesssim 1$, we calculate the upper limits (ULs) on projected mass ($m\sin{i}$) of the FFO as a function of periapsis distance ($p$).
To do this, we first divide the range for the values of $\log{p}$ (AU) into separate bins with bin widths of $\sim 1$.
Thereafter, we calculate the 95\% UL on $\log(m \sin{i})$ for each $\log{p}$ bin from the posterior distribution.
As the prior for $m \sin{i}$ is log-uniform, we use reweighting method to calculate the expected posterior distribution for $\log(m \sin{i})$ for a uniform prior in $m \sin{i}$ while calculating the ULs.
This ensures that the calculated ULs are robust and do not depend on the lower bound of the $m \sin{i}$ prior \citep{HourihaneMeyers+2022}.

\section{Results}
\label{sec:results}

In this section, we present the findings of our search for a hyperbolic gravitational scattering event of an FFO for all 68 pulsars in the NANOGrav 15-year dataset.
As a scattering event involves a single pulsar and an FFO, our search is done for one pulsar at a time.
We did not detect any statistically significant event in our dataset as the SD Bayes factor $\mathcal{B} \lesssim 1$ for every pulsar.
Therefore, we calculated ULs on $\log{(m \sin{i})}$ for various ranges of periapsis distance for each pulsar.

In Figure~\ref{fig:posterior_ULs}, we present illustrative results for PSR J0030+0451.
The top panel of Figure~\ref{fig:posterior_ULs} shows the corner plot of the model parameters for the hyperbolic scattering event with 1D and 2D posterior distributions of the parameters.
As evident from the 2-D posterior distribution of $\log{(m \sin{i})}-\log{p}$, we see that some part of the parameter space (towards large $m\sin{i}$ and small $p$) is ruled out by the data.

In the bottom panel, we present a violin plot showing the binned posterior distributions of the FFO projected mass as a function of $\log{p}$.
The 95\% ULs on the projected mass for different $\log{p}$ bins are also shown with horizontal blue dashes.
In other words, for a particular $\log{p}$ bin, we can say with 95\% confidence that no FFO with projected mass greater than the UL value has passed by the pulsar within the time-span of our dataset with distance of closest approach in the range of that given bin.
Note that for a few bins with large $\log{p}$ values, we do not show any ULs. 
This is because the ULs on the projected mass for those bins is restricted by the prior and not informed by the data and therefore not physically meaningful.
If the calculated 95\% UL value of $\log{(m \sin{i})}$ is more than $-1.2$, then we consider that UL is constrained by the prior rather than the data (see \citealt{NG12p5_2024_3C66B_ecc} for details).
In Table~\ref{tab:mass_ULs}, we show the 95\% ULs on the projected mass of the FFO for different bins of $\log{p}$ for all 68 pulsars.

\begingroup

\setlength{\tabcolsep}{12pt}
\startlongtable
\begin{deluxetable*}{l c c c c c c}
\tablecaption{95\% ULs on the $\log{(m \sin{i})}$ of FFOs for different $\log{p}$ bins. 
The dash sign ($-$) represents that the UL for that bin is either not available or not physically meaningful due to the prior for the model parameters.
\label{tab:mass_ULs}}
\tablehead{ & \multicolumn{6}{c}{ULs on $\log{(m \sin{i}/M_{\odot})}$ for $\log{( p/\text{AU})}$ bins around} \\[0pt] \cline{2-7}
    \colhead{PSR} &
    \colhead{$-2$} &
    \colhead{$-1$} &
    \colhead{$0$} &
    \colhead{$1$} & 
    \colhead{$2$} &
    \colhead{$3$}
    }
\startdata
B1855+09 & $-$ & $-$ & $-7.70$ & $-7.22$ & $-4.46$ & $-1.40$\\
B1937+21 & $-7.83$ & $-7.92$ & $-7.60$ & $-6.09$ & $-4.10$ & $-1.29$\\
B1953+29 & $-$ & $-$ & $-6.70$ & $-6.49$ & $-3.26$ & $-$\\
J0023+0923 & $-$ & $-8.32$ & $-7.72$ & $-7.13$ & $-4.36$ & $-$\\
J0030+0451 & $-7.99$ & $-8.37$ & $-8.08$ & $-7.02$ & $-4.02$ & $-$\\
J0340+4130 & $-8.06$ & $-7.78$ & $-7.62$ & $-6.72$ & $-3.53$ & $-1.26$\\
J0406+3039 & $-$ & $-$ & $-7.21$ & $-5.82$ & $-2.25$ & $-$\\
J0437$-$4715 & $-$ & $-$ & $-7.45$ & $-6.44$ & $-3.30$ & $-$\\
J0509+0856 & $-$ & $-$ & $-6.30$ & $-4.46$ & $-1.30$ & $-$\\
J0557+1551 & $-$ & $-$ & $-6.52$ & $-5.75$ & $-2.47$ & $-$\\
J0605+3757 & $-$ & $-$ & $-6.78$ & $-5.02$ & $-1.81$ & $-$\\
J0610$-$2100 & $-$ & $-7.63$ & $-7.22$ & $-5.38$ & $-2.16$ & $-$\\
J0613$-$0200 & $-$ & $-8.46$ & $-8.07$ & $-7.32$ & $-4.86$ & $-1.56$\\
J0614$-$3329 & $-$ & $-$ & $-6.84$ & $-4.73$ & $-1.25$ & $-$\\
J0636+5128 & $-$ & $-8.11$ & $-7.89$ & $-7.01$ & $-4.01$ & $-$\\
J0645+5158 & $-7.97$ & $-8.08$ & $-7.48$ & $-6.80$ & $-4.56$ & $-1.33$\\
J0709+0458 & $-$ & $-$ & $-6.69$ & $-5.52$ & $-1.68$ & $-$\\
J0740+6620 & $-$ & $-$ & $-8.16$ & $-7.53$ & $-3.85$ & $-1.42$\\
J0931$-$1902 & $-7.68$ & $-7.27$ & $-7.53$ & $-6.76$ & $-3.23$ & $-1.31$\\
J1012+5307 & $-$ & $-7.81$ & $-7.47$ & $-6.77$ & $-3.87$ & $-1.53$\\
J1012$-$4235 & $-$ & $-$ & $-6.31$ & $-4.73$ & $-1.68$ & $-$\\
J1022+1001 & $-$ & $-$ & $-6.23$ & $-5.98$ & $-2.92$ & $-$\\
J1024$-$0719 & $-7.95$ & $-8.30$ & $-7.74$ & $-6.59$ & $-1.51$ & $-$\\
J1125+7819 & $-$ & $-$ & $-7.46$ & $-6.72$ & $-3.38$ & $-$\\
J1312+0051 & $-$ & $-$ & $-6.82$ & $-5.53$ & $-2.74$ & $-$\\
J1453+1902 & $-7.25$ & $-6.84$ & $-7.14$ & $-6.61$ & $-3.60$ & $-$\\
J1455$-$3330 & $-$ & $-$ & $-7.09$ & $-7.35$ & $-4.03$ & $-1.56$\\
J1600$-$3053 & $-$ & $-$ & $-8.27$ & $-7.08$ & $-4.88$ & $-1.52$\\
J1614$-$2230 & $-$ & $-$ & $-8.40$ & $-7.49$ & $-4.55$ & $-1.28$\\
J1630+3734 & $-$ & $-$ & $-7.01$ & $-5.75$ & $-2.10$ & $-$\\
J1640+2224 & $-$ & $-$ & $-8.03$ & $-7.47$ & $-5.12$ & $-1.54$\\
J1643$-$1224 & $-$ & $-$ & $-$ & $-6.19$ & $-4.22$ & $-1.49$\\
J1705$-$1903 & $-$ & $-7.14$ & $-7.10$ & $-5.30$ & $-2.09$ & $-$\\
J1713+0747 & $-$ & $-$ & $-8.38$ & $-7.86$ & $-5.29$ & $-1.96$\\
J1719$-$1438 & $-7.26$ & $-7.45$ & $-7.23$ & $-5.94$ & $-1.86$ & $-$\\
J1730$-$2304 & $-7.77$ & $-7.86$ & $-6.84$ & $-5.72$ & $-2.60$ & $-$\\
J1738+0333 & $-$ & $-8.24$ & $-7.21$ & $-6.82$ & $-4.02$ & $-1.25$\\
J1741+1351 & $-$ & $-$ & $-8.19$ & $-7.43$ & $-4.07$ & $-$\\
J1744$-$1134 & $-7.03$ & $-7.28$ & $-6.47$ & $-6.89$ & $-4.31$ & $-1.54$\\
J1745+1017 & $-$ & $-6.75$ & $-6.65$ & $-5.01$ & $-1.79$ & $-$\\
J1747$-$4036 & $-7.20$ & $-7.55$ & $-7.00$ & $-6.05$ & $-3.13$ & $-$\\
J1751$-$2857 & $-$ & $-$ & $-6.73$ & $-5.81$ & $-2.09$ & $-$\\
J1802$-$2124 & $-$ & $-6.63$ & $-6.70$ & $-5.42$ & $-2.02$ & $-$\\
J1811$-$2405 & $-$ & $-$ & $-7.03$ & $-6.52$ & $-3.03$ & $-$\\
J1832$-$0836 & $-7.07$ & $-7.21$ & $-7.24$ & $-7.16$ & $-3.69$ & $-$\\
J1843$-$1113 & $-7.94$ & $-7.44$ & $-7.22$ & $-6.22$ & $-2.11$ & $-$\\
J1853+1303 & $-$ & $-$ & $-$ & $-7.06$ & $-3.72$ & $-1.31$\\
J1903+0327 & $-$ & $-$ & $-7.05$ & $-6.10$ & $-2.95$ & $-1.26$\\
J1909$-$3744 & $-$ & $-$ & $-8.52$ & $-7.68$ & $-5.01$ & $-$\\
J1910+1256 & $-$ & $-$ & $-7.94$ & $-7.31$ & $-4.04$ & $-1.52$\\
J1911+1347 & $-8.69$ & $-8.45$ & $-7.32$ & $-7.44$ & $-4.28$ & $-$\\
J1918$-$0642 & $-$ & $-$ & $-7.41$ & $-7.02$ & $-4.90$ & $-1.27$\\
J1923+2515 & $-8.00$ & $-8.21$ & $-7.79$ & $-6.98$ & $-4.02$ & $-$\\
J1944+0907 & $-7.86$ & $-7.02$ & $-7.51$ & $-7.02$ & $-4.45$ & $-$\\
J1946+3417 & $-$ & $-$ & $-7.41$ & $-6.25$ & $-3.19$ & $-$\\
J2010$-$1323 & $-8.11$ & $-7.97$ & $-7.72$ & $-7.46$ & $-4.79$ & $-$\\
J2017+0603 & $-$ & $-$ & $-7.53$ & $-6.89$ & $-3.89$ & $-1.34$\\
J2033+1734 & $-$ & $-$ & $-6.85$ & $-6.73$ & $-2.77$ & $-1.27$\\
J2043+1711 & $-$ & $-8.77$ & $-7.84$ & $-7.38$ & $-4.89$ & $-1.50$\\
J2124$-$3358 & $-7.37$ & $-7.40$ & $-7.00$ & $-6.07$ & $-2.34$ & $-$\\
J2145$-$0750 & $-$ & $-$ & $-7.59$ & $-6.62$ & $-4.80$ & $-1.58$\\
J2214+3000 & $-$ & $-8.01$ & $-7.00$ & $-6.70$ & $-3.53$ & $-1.26$\\
J2229+2643 & $-$ & $-$ & $-7.45$ & $-7.10$ & $-3.93$ & $-1.28$\\
J2234+0611 & $-$ & $-$ & $-8.04$ & $-6.62$ & $-2.88$ & $-$\\
J2234+0944 & $-$ & $-8.02$ & $-7.55$ & $-7.07$ & $-3.68$ & $-$\\
J2302+4442 & $-$ & $-$ & $-$ & $-6.41$ & $-3.19$ & $-$\\
J2317+1439 & $-$ & $-$ & $-7.83$ & $-7.73$ & $-4.70$ & $-1.91$\\
J2322+2057 & $-8.05$ & $-7.31$ & $-7.95$ & $-7.02$ & $-3.53$ & $-1.30$\\
\enddata
\end{deluxetable*}

\endgroup

\subsection{Insights into FFO Population}
\label{subsec:FFO_density}

Although we did not detect any gravitational scattering events of FFOs by a pulsar in our data, we utilize the posterior distributions of the model parameters to impose constraints on the population characteristics of FFOs in the vicinity of the pulsars present in the dataset.
To accomplish this, we first employ Equation~\eqref{eqn:b_p_relation} to compute the values of the impact parameter $b$ corresponding to each sample from the posterior distribution.
It is evident that a portion of the 2D posterior distribution of $\log{(m \sin{i})}-\log{b}$ is also ruled out by the data, in a manner analogous to the 2D posterior distribution of $\log{(m \sin{i})}-\log{p}$ as observed in the top panel of Figure~\ref{fig:posterior_ULs}.

Thereafter, we divide the samples into different bins for $\log{(m \sin{i})}$ centered around integer values with bin width of 1.
Now, for a specific FFO mass bin, if the minimum allowed value of $b$ by the timing data of a pulsar is $b_{\rm min}$, then it can be shown that the approximate FFO number density in the region close to that pulsar for that mass range
\begin{align}
    n_{\rm FFO} \lesssim \frac{1}{\pi\, b_{\rm min}^2 \,v_{\rm avg.}\, T_{\rm span}} \,, 
    \label{eq:nFFO}
\end{align}
where $v_{\rm avg.} \approx 87.77$ km/s is the mean of the prior for $v_{\infty}$, and $T_{\rm span}$ is the total time span of the pulsar data (see Appendix~\ref{app:nFFO_derivation} for more details). 
From this, we can calculate the 95\% ULs on $n_{\rm FFO}$ for that particular FFO mass range as
\begin{align}
    \left. n_{\rm FFO}\right|_{\rm UL}^{95\%} = \frac{1}{\pi\,v_{\rm avg.}\,T_{\rm span}} \left. \frac{1}{b^2}\right|_{\rm UL}^{95\%}\,,
    \label{eq:nFFO_UL95}
\end{align}
where $\left. 1/b^2\right|^{95\%}_{\rm UL}$ represents the 95\% UL on $(1/b^2)$ for that mass bin.
To ensure that the effective prior distribution for $n_{\rm FFO}$ is uniform when estimating the UL, we apply the prior reweighting method.

Further, if we consider the FFO number density to be approximately constant throughout the Galaxy, we can combine the results from multiple pulsars to calculate a more stringent constraint on $n_{\rm FFO}$ using the following formula,
\begin{align}
    \left. n_{\rm FFO}\right|_{\rm UL,\,combined}^{95\%} = \frac{1}{\pi\,v_{\rm avg.} \sum\limits_{psr} \left(T_{\text{span},\,psr} \,\,b^2|_{\text{min},\,psr}^{95\%}\right)}\,,
    \label{eq:nFFO_UL95_combined}
\end{align}
where $b^2|_{\text{min},\,psr}^{95\%}$ is defined as
\begin{align}
    b^2|_{\text{min},\,psr}^{95\%} = \left( \left. \frac{1}{b^2}\right|_{\text{UL},\,psr}^{95\%}\right)^{-1}\,,
\end{align}
and the summation is over all the pulsars. This stronger constraint replaces the volume sampled by one pulsar with the total volume sampled by all of the pulsars in the array (see Appendix \ref{app:nFFO_derivation}).

\begin{figure*}
    \centering
    \includegraphics[width=0.8\linewidth]{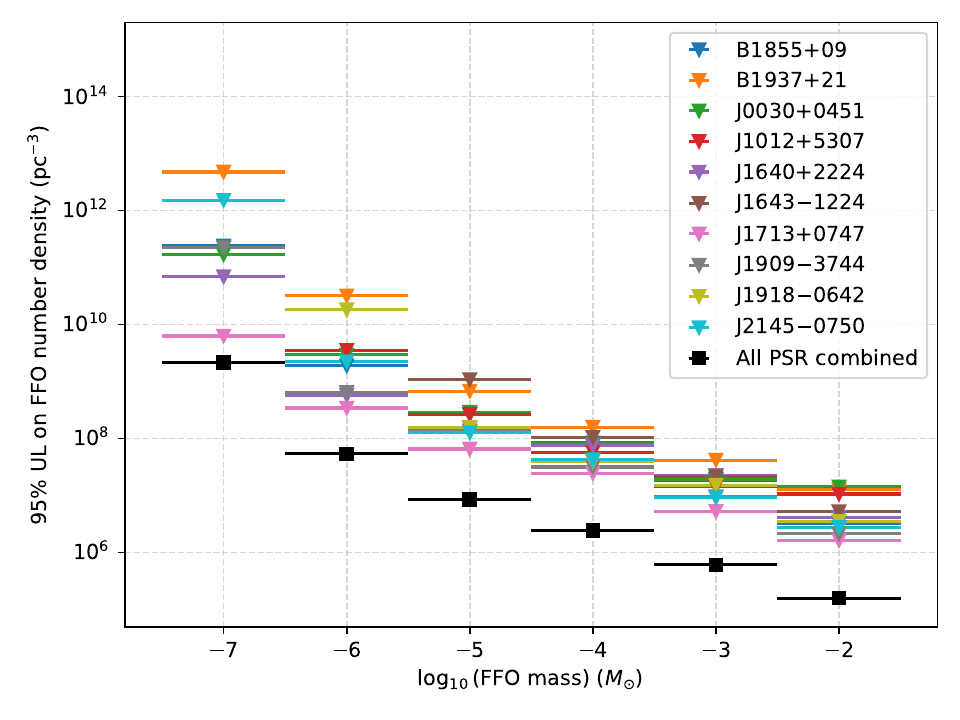}
    \caption{95\% ULs on the number density of FFOs for different mass ranges in the vicinity of 10 of the longest timed pulsars in the NG15 dataset. 
    The combined upper limits on the FFO number density from all 68 pulsars are also shown with black squares.
    }
    \label{fig:nFFO_ULs}
\end{figure*}

In Figure~\ref{fig:nFFO_ULs}, we present the 95\% ULs on the FFO number density across different mass ranges, derived from 10 of the longest timed pulsars in the NG15 dataset. 
For mass bins where the UL on $n_{\rm FFO}$ is constrained solely by the prior rather than by the data, we omit the corresponding results from the figure, as they are not physically meaningful.
Further, the combined ULs on $n_{\rm FFO}$, computed using Equation~\eqref{eq:nFFO_UL95_combined} and considering all 68 pulsars, are also presented for each FFO mass bin in Figure~\ref{fig:nFFO_ULs}.

It is important to note that the ULs presented here for the FFO mass bins correspond to ranges in $\log{(m \sin{i})}$ rather than $\log{m}$ itself. 
Given that the statistical average of $\sin{i}$ for uniformly distributed viewing geometries with $i$ ranging from $0$ to $\pi$ radians is $\pi/4$, one can correct for this by subtracting $\log{(\pi/4)}$ ($\approx -0.1$) from the $\log{(m \sin{i})}$ values. 
This adjustment yields the corresponding ULs for average FFO mass bins, instead of projected mass. 
However, since these ULs are derived for broad, order-of-magnitude bins of FFO mass, we have neglected the $\sin{i}$ correction when presenting the results.

In Table~\ref{tab:nFFO_ULs}, we have given the 95\% ULs on $\log{n_{\rm FFO}}$ for different mass ranges of the FFO in the vicinity of all 68 pulsars in the NG15 dataset.
For mass bins where the calculated UL is constrained primarily by the prior distribution and thus not physically meaningful, we indicate the result with a dash symbol ($-$).

\begingroup

\setlength{\tabcolsep}{18pt}
\startlongtable
\begin{deluxetable*}{l c c c c c c}
\tablecaption{95\% ULs on $\log{n_{\rm FFO}}$  (pc$^{-3}$) for different $\log{(m/M_{\odot})}$ bins. \label{tab:nFFO_ULs}}
\tablehead{ & \multicolumn{6}{c}{ $\log{(m/M_{\odot})}$ bin centers} \\[0pt] \cline{2-7}
    \colhead{PSR} &
    \colhead{$-7$} &
    \colhead{$-6$} &
    \colhead{$-5$} & 
    \colhead{$-4$} &
    \colhead{$-3$} &
    \colhead{$-2$}
    }
\startdata
All PSRs & $9.33$ & $7.73$ & $6.93$ & $6.39$ & $5.78$ & $5.20$  \\
\hline
B1855+09 & $11.38$ & $9.28$ & $8.43$ & $7.92$ & $7.27$ & $6.51$\\
B1937+21 & $12.68$ & $10.50$ & $8.83$ & $8.20$ & $7.62$ & $7.10$\\
B1953+29 & $-$ & $11.24$ & $9.11$ & $8.39$ & $8.03$ & $7.23$\\
J0023+0923 & $-$ & $9.31$ & $8.70$ & $8.03$ & $7.40$ & $7.30$\\
J0030+0451 & $11.23$ & $9.47$ & $8.46$ & $7.91$ & $7.29$ & $7.15$\\
J0340+4130 & $12.35$ & $9.86$ & $9.02$ & $8.44$ & $8.01$ & $7.08$\\
J0406+3039 & $-$ & $11.38$ & $10.34$ & $9.86$ & $9.29$ & $8.61$\\
J0437$-$4715 & $-$ & $11.31$ & $9.61$ & $8.99$ & $8.57$ & $7.88$\\
J0509+0856 & $-$ & $-$ & $11.44$ & $10.19$ & $9.59$ & $8.95$\\
J0557+1551 & $-$ & $-$ & $10.42$ & $9.60$ & $9.09$ & $8.48$\\
J0605+3757 & $-$ & $-$ & $11.08$ & $10.11$ & $9.56$ & $8.88$\\
J0610$-$2100 & $-$ & $11.57$ & $10.72$ & $10.20$ & $9.58$ & $9.06$\\
J0613$-$0200 & $10.43$ & $9.18$ & $8.37$ & $7.70$ & $6.99$ & $6.56$\\
J0614$-$3329 & $-$ & $12.36$ & $11.04$ & $10.38$ & $9.73$ & $9.21$\\
J0636+5128 & $13.72$ & $10.16$ & $9.36$ & $8.68$ & $8.00$ & $7.76$\\
J0645+5158 & $13.25$ & $9.60$ & $8.71$ & $8.03$ & $8.06$ & $7.12$\\
J0709+0458 & $-$ & $12.49$ & $10.44$ & $9.69$ & $9.23$ & $8.67$\\
J0740+6620 & $10.63$ & $9.62$ & $9.03$ & $8.23$ & $7.75$ & $7.20$\\
J0931$-$1902 & $-$ & $10.07$ & $9.63$ & $8.64$ & $8.05$ & $7.38$\\
J1012+5307 & $-$ & $9.55$ & $8.42$ & $7.76$ & $7.16$ & $7.03$\\
J1012$-$4235 & $-$ & $-$ & $10.83$ & $10.20$ & $9.38$ & $9.07$\\
J1022+1001 & $-$ & $-$ & $9.96$ & $9.27$ & $8.85$ & $8.23$\\
J1024$-$0719 & $12.14$ & $9.95$ & $9.47$ & $8.99$ & $8.73$ & $8.45$\\
J1125+7819 & $-$ & $10.07$ & $9.40$ & $8.85$ & $8.37$ & $7.55$\\
J1312+0051 & $-$ & $11.93$ & $10.17$ & $9.55$ & $8.84$ & $8.33$\\
J1453+1902 & $-$ & $10.09$ & $9.25$ & $8.80$ & $8.24$ & $7.77$\\
J1455$-$3330 & $-$ & $9.18$ & $8.42$ & $8.28$ & $7.57$ & $6.62$\\
J1600$-$3053 & $10.47$ & $8.96$ & $8.27$ & $7.85$ & $7.01$ & $6.35$\\
J1614$-$2230 & $10.87$ & $9.08$ & $8.49$ & $7.97$ & $7.70$ & $7.02$\\
J1630+3734 & $-$ & $12.04$ & $10.32$ & $9.73$ & $9.28$ & $8.62$\\
J1640+2224 & $10.84$ & $8.76$ & $8.18$ & $7.88$ & $7.35$ & $6.62$\\
J1643$-$1224 & $-$ & $-$ & $9.04$ & $8.02$ & $7.33$ & $6.72$\\
J1705$-$1903 & $-$ & $11.06$ & $10.44$ & $9.75$ & $8.89$ & $9.05$\\
J1713+0747 & $9.80$ & $8.54$ & $7.82$ & $7.39$ & $6.72$ & $6.21$\\
J1719$-$1438 & $-$ & $12.04$ & $10.32$ & $9.78$ & $9.25$ & $8.72$\\
J1730$-$2304 & $14.11$ & $10.94$ & $10.27$ & $9.58$ & $9.39$ & $8.28$\\
J1738+0333 & $13.18$ & $9.86$ & $9.01$ & $8.30$ & $8.02$ & $7.24$\\
J1741+1351 & $11.06$ & $9.28$ & $8.53$ & $8.08$ & $7.44$ & $6.83$\\
J1744$-$1134 & $-$ & $12.95$ & $8.46$ & $7.92$ & $7.17$ & $6.89$\\
J1745+1017 & $-$ & $12.98$ & $10.61$ & $9.94$ & $9.28$ & $8.86$\\
J1747$-$4036 & $-$ & $11.64$ & $9.86$ & $9.06$ & $8.35$ & $7.86$\\
J1751$-$2857 & $-$ & $11.50$ & $10.39$ & $9.73$ & $9.47$ & $8.61$\\
J1802$-$2124 & $-$ & $13.12$ & $10.67$ & $10.15$ & $9.55$ & $8.98$\\
J1811$-$2405 & $12.92$ & $10.52$ & $10.05$ & $9.50$ & $8.64$ & $8.35$\\
J1832$-$0836 & $-$ & $9.68$ & $9.04$ & $8.52$ & $8.06$ & $7.28$\\
J1843$-$1113 & $14.76$ & $10.76$ & $10.12$ & $9.41$ & $8.84$ & $8.83$\\
J1853+1303 & $-$ & $9.49$ & $8.94$ & $8.28$ & $7.64$ & $7.06$\\
J1903+0327 & $-$ & $-$ & $9.94$ & $8.57$ & $7.95$ & $7.56$\\
J1909$-$3744 & $11.35$ & $8.81$ & $8.13$ & $7.50$ & $6.98$ & $6.33$\\
J1910+1256 & $11.71$ & $9.44$ & $8.73$ & $8.33$ & $7.67$ & $7.21$\\
J1911+1347 & $12.97$ & $9.48$ & $8.83$ & $8.29$ & $7.91$ & $7.31$\\
J1918$-$0642 & $-$ & $10.26$ & $8.20$ & $7.60$ & $7.18$ & $6.54$\\
J1923+2515 & $12.73$ & $9.48$ & $8.88$ & $8.35$ & $7.63$ & $7.09$\\
J1944+0907 & $14.79$ & $10.01$ & $8.96$ & $8.06$ & $7.49$ & $6.79$\\
J1946+3417 & $-$ & $10.89$ & $9.73$ & $9.03$ & $8.58$ & $7.91$\\
J2010$-$1323 & $11.60$ & $9.22$ & $8.82$ & $8.06$ & $7.45$ & $6.86$\\
J2017+0603 & $-$ & $9.44$ & $8.82$ & $8.42$ & $7.82$ & $6.98$\\
J2033+1734 & $-$ & $10.57$ & $9.16$ & $8.58$ & $8.25$ & $7.33$\\
J2043+1711 & $11.06$ & $9.09$ & $8.47$ & $7.89$ & $7.32$ & $6.81$\\
J2124$-$3358 & $-$ & $11.60$ & $10.16$ & $9.59$ & $9.12$ & $8.46$\\
J2145$-$0750 & $12.18$ & $9.35$ & $8.10$ & $7.63$ & $6.97$ & $6.44$\\
J2214+3000 & $13.98$ & $9.63$ & $8.99$ & $8.39$ & $8.15$ & $7.30$\\
J2229+2643 & $-$ & $9.66$ & $9.04$ & $8.58$ & $8.04$ & $7.32$\\
J2234+0611 & $11.40$ & $10.32$ & $9.29$ & $8.49$ & $8.15$ & $7.61$\\
J2234+0944 & $12.65$ & $9.85$ & $9.22$ & $8.57$ & $7.91$ & $7.30$\\
J2302+4442 & $-$ & $10.87$ & $9.16$ & $8.49$ & $7.90$ & $7.17$\\
J2317+1439 & $9.86$ & $9.04$ & $8.18$ & $7.80$ & $7.00$ & $6.35$\\
J2322+2057 & $15.28$ & $9.99$ & $9.33$ & $8.75$ & $8.22$ & $7.80$\\
\enddata
\end{deluxetable*}

\endgroup

\section{Summary and Discussion}
\label{sec:summary}

In this work, we utilized the NANOGrav 15-year narrow-band timing dataset to search for hyperbolic gravitational scattering events of FFOs by pulsars in the Milky Way. 
While no such events were detected, our analysis allowed us to constrain part of the relevant parameter space. 
Based on these constraints, we derived upper limits on the FFO number density in the vicinity of pulsars across a range of FFO masses.
Furthermore, by combining results from all 68 pulsars, we obtained a significantly stronger constraint on the FFO number density.


We find that the upper limits on FFO number densities derived from PTA data, although consistent with expectations, are several order of magnitude higher than those predicted by simulations or inferred from microlensing surveys for the planetary mass range \citep{Coleman+2025_FFP_polulation_simulation, Sumi+2023_FFP_mass_function, Hadden&WU_2025_FFP_neptune_mass}. 
Furthermore, the current estimated upper limits on FFO number densities for different mass ranges correspond to mass density constraints spanning approximately $50 - 1600 \text{M}_{\odot}/\text{pc}^3$. 
These limits are also orders of magnitude higher than the expected dark matter density in the Milky Way \citep[approximately $0.01\text{M}_{\odot}\text{pc}^{-3}$,][]{deSalas+2021_rho_DM}.
Consequently, our results do not exclude any mass range for PBHs as potential dark matter candidates, consistent with the findings of \citet{NG2023_15yr_new_physics}. 
In contrast, microlensing surveys have already ruled out PBHs and other compact objects with masses ranging from $10^{-8}$ to $10^3\text{M}_{\odot}$ as dominant components of dark matter \citep{Mroz+2025_PBH_DM_rule_out}.

However, this paper presents a new approach of using pulsars to search for FFOs in the interstellar space and to constrain their number density in the vicinity of pulsars in PTA datasets, which are distributed in our local region of the Galaxy.
The mass range of FFOs accessible to this method is also broad, thus this approach can potentially complement other methods such as microlensing in near future.
From Equation~\eqref{eq:nFFO_UL95}, we see that $\left. n_{\rm FFO}\right|_{\rm UL}^{95\%} \propto 1/T_\mathrm{span}$. 
Therefore, as PTA experiments continue to monitor MSPs with increasing time baselines, the sensitivity to detect gravitational perturbations from passing FFOs will improve, enabling even tighter constraints on their number density \citep[see Figure 6 of][for example]{Jennings_2020_ISO_Psr_scattering}.
Furthermore, the increasing time baselines will increase the probability of a positive FFO detection, as the volume of the Galaxy sampled increases linearly with time (see Equation~\eqref{eq:volume}).

Future datasets from advanced radio observatories such as the Five-hundred-meter Aperture Spherical Telescope, DSA-2000, and the Square Kilometer Array, and use of robust TOA generation techniques \citep[e.g.,][]{Rogers+2024, Dey+2024_GBT_pol_cal}, are expected to deliver improved timing precision, further allowing us to put more stringent ULs on the FFO number density as well as increasing the chance of detection. 
The upcoming third data release from the International Pulsar Timing Array collaboration \citep{Hobbs+2010_IPTA}, which will combine data from multiple PTA collaborations worldwide, will include a larger pulsar sample and longer observation spans. 
This expanded dataset promises a valuable opportunity to detect hyperbolic FFO–pulsar encounters and explore the underlying population properties of FFOs in our Galaxy.

\section*{Data availability}
The NANOGrav 15-year narrowband data set is available at \url{https://zenodo.org/records/14773896}.

\section*{Author Contributions}
L.D. led the project, carried the analyses, and prepared the text, figures, and tables with inputs from R.J.J., J.D.T., J.G., M.A.M., and T.J.W.L..
J.D.T. carried out the three body simulations and prepared the text for the relevant part of the paper with input from J.G.
The rest of the authors (including R.J.J., J.G., and M.A.M.) contributed to the collection and analysis of the NANOGrav 15 year dataset.

\section*{Acknowledgements}

This work has been carried out as part of the NANOGrav collaboration, which receives support from the National Science Foundation (NSF) Physics Frontiers Center award numbers 1430284 and 2020265. 
The Arecibo Observatory is a facility of the NSF operated under a cooperative agreement (No. AST-1744119) by the University of Central Florida (UCF) in alliance with Universidad Ana G. M{\'e}ndez (UAGM) and Yang Enterprises (YEI), Inc. 
The Green Bank Observatory is a facility of the NSF operated under a cooperative agreement by Associated Universities, Inc. 
The National Radio Astronomy Observatory is a facility of the NSF operated under a cooperative agreement by Associated Universities, Inc.
Computational resources were provided by the Link HPC cluster and cyber-infrastructure, which is maintained by the Center for Gravitational Waves and Cosmology at West Virginia University and funded in part by NSF IIA-1458952 \& NSF PHY-2020265.
L.D. is supported by a West Virginia University postdoctoral fellowship.
J.D.T. is supported by the NASA West Virginia Space Grant Consortium, Grant~\#~80NSSC25M7079.
P.R.B.\ is supported by the Science and Technology Facilities Council, grant number ST/W000946/1.
Pulsar research at UBC is supported by an NSERC Discovery Grant and by CIFAR.
K.C.\ is supported by a UBC Four Year Fellowship (6456).
M.E.D.\ acknowledges support from the Naval Research Laboratory by NASA under contract S-15633Y.
T.D.\ and M.T.L.\ received support by an NSF Astronomy and Astrophysics Grant (AAG) award number 2009468 during this work.
E.C.F.\ is supported by NASA under award number 80GSFC24M0006.
G.E.F.\ is supported by NSF award PHY-2011772.
D.R.L.\ and M.A.M.\ are supported by NSF \#1458952.
M.A.M.\ is supported by NSF \#2009425.
The Dunlap Institute is funded by an endowment established by the David Dunlap family and the University of Toronto.
T.T.P.\ acknowledges support from the Extragalactic Astrophysics Research Group at E\"{o}tv\"{o}s Lor\'{a}nd University, funded by the E\"{o}tv\"{o}s Lor\'{a}nd Research Network (ELKH), which was used during the development of this research.
H.A.R.\ is supported by NSF Partnerships for Research and Education in Physics (PREP) award No.\ 2216793.
S.M.R.\ and I.H.S.\ are CIFAR Fellows.
Portions of this work performed at NRL were supported by ONR 6.1 basic research funding.

\software{
\texttt{ENTERPRISE} \citep{EllisVallisneri+2017,JohnsonMeyers+2023},
\texttt{enterprise\_extensions} \citep{TaylorBaker+2021,JohnsonMeyers+2023},
\texttt{PTMCMCSampler} \citep{EllisvanHaasteren2017,JohnsonMeyers+2023},
\texttt{numpy} \citep{HarrisMillman+2020}, 
\texttt{matplotlib} \citep{Hunter2007}, 
\texttt{corner} \citep{Foreman-Mackey2016},
\texttt{rebound} \citep{rebound, reboundias15}.
}

\appendix
\section{Calculating upper limits on FFO number density}
\label{app:nFFO_derivation}

Assuming the same velocity $v_{\infty}$ for all FFOs, the volume containing FFOs that will pass the pulsar at an impact parameter less than $b_{\text{min}}$ within the span of the data set $T_{\text{span}}$ is $V = \pi\,b_{\text{min}}^2\,v_{\infty}\,T_{\text{span}}$ \citep{Jennings_2020_ISO_Psr_scattering}.
Now averaging $v_{\infty}$ over the prior range, the velocity-averaged volume is
\begin{align}
    V &= \pi\,b_{\text{min}}^2\,v_{\text{avg.}}\,T_{\text{span}}\,.
    \label{eq:volume}
\end{align}
As no hyperbolic gravitational scattering event of an FFO is detected, the minimum allowed value for the impact parameter $b_{\text{min}}$ implies that the number of FFOs within the volume $V$ is $\lesssim 1$. Therefore, the number density of FFOs in the vicinity of the pulsar is
\begin{align}
    n_{\text{FFO}} \lesssim \frac{1}{\pi\,b_{\text{min}}^2\,v_{\text{avg.}}\,T_{\text{span}}} \,.
\end{align}

Now, let us assume a PTA with $N_{psr}$ pulsars with minimum allowed impact parameters of $b_{\text{min},\,psr}$ and data spans of $T_{\text{span},\,psr}$.
Combining the limits from all the pulsars, the total volume within which the number of FFOs is $\lesssim 1$ is given by
\begin{align}
    V_{\text{combined}} = \pi\,v_{\text{avg.}}\,\sum\limits_{psr} \left(b_{\text{min},\,psr}^2\, T_{\text{span},\,psr}\right) \,.
\end{align}
The resulting combined upper limit on $n_{\text{FFO}}$ is therefore given by
\begin{align}
    n_{\text{FFO,\,combined}} \lesssim \frac{1}{\pi\,v_{\rm avg.} \sum\limits_{psr} \left(T_{\text{span},\,psr} \,\,b^2_{\text{min},\,psr}\right)}\,.
\end{align}

\bibliography{references}{}
\bibliographystyle{aasjournal}

\end{document}